\documentclass[12pt,tightenlines,showpacs]{revtex4-1}
\usepackage{graphics,graphicx,epsfig,amsmath,amssymb,amsfonts}
\usepackage{color}
\begin{document}
\title{Scaling relations in the diffusive infiltration in fractals}
\author{F. D. A. Aar\~ao Reis\footnote{Email address: reis@if.uff.br}}
\affiliation{Instituto de F\'\i sica, Universidade Federal Fluminense,\\
Avenida Litor\^anea s/n, 24210-340 Niter\'oi RJ, Brazil}
\date{\today}

\begin{abstract}

In a recent work on fluid infiltration in a Hele-Shaw cell with the pore-block geometry of
Sierpinski carpets (SCs), the area filled by the invading fluid was shown to scale as $F\sim t^n$,
with $n<1/2$, thus providing a macroscopic realization of anomalous diffusion
[Filipovitch et al, Water Resour. Res. {\bf 52} 5167 (2016)].
The results agree with simulations of a diffusion equation with constant pressure at one of the
borders of those fractals, but the exponent $n$ is very different from the anomalous exponent
$\nu=1/D_W$ of single particle diffusion in the same fractals ($D_W$ is the random walk dimension).
Here we use a scaling approach to show that those exponents are related as
$n=\nu\left( D_F-D_B\right)$, where $D_F$ and $D_B$ are the fractal dimensions of the bulk and
of the border from which diffusing particles come, respectively.
This relation is supported by accurate numerical estimates in two SCs and in two generalized
Menger sponges (MSs), in which we performed simulations of single particle random walks (RWs)
with a rigid impermeable border and of a diffusive infiltration model in which
that border is permanently filled with diffusing particles.
This study includes one MS whose external border is also fractal.
The exponent relation is also consistent with the recent simulational and experimental results on
fluid infiltration in SCs, and explains the approximate quadratic dependence of $n$ on $D_F$ in
these fractals.
We also show that the mean-square displacement of single particle RWs has log-periodic oscillations,
whose periods are similar for fractals with the same scaling factor in the generator (even with
different embedding dimensions), which is consistent with the discrete scale invariance scenario.
The roughness of a diffusion front defined in the infiltration problem also shows this type of
oscillation, which is enhanced in fractals with narrow channels between large lacunas.

\end{abstract}

\maketitle

\section{Introduction}
\label{intro}

Anomalous diffusion is frequently observed in transport in porous media and was subject of
intense theoretical research in recent decades \cite{bouchaud,havlin,metzler2000,metzler2014,xu}.
In such media, the mean-square displacement $R$ of a tracer particle scales in time $t$ as
\begin{equation}
R \sim t^\nu ,
\label{R}
\end{equation}
where, in the case of subdiffusion, $\nu<1/2$.
The random walk dimension
\begin{equation}
D_W=\frac{1}{\nu}
\label{DW}
\end{equation}
is consequently larger than $2$.
In normal or Fickean diffusion, $\nu =1/2$ ($D_W=2$); superdiffusion is characterized by
$\nu>1/2$, but such case is not discussed here.
The delay in material transport in a porous medium is caused by irregularities such as impenetrable
barriers and dead ends.
The anomaly is observed if this delay has no characteristic time scale \cite{havlin},
which is in turn related to the absence of a characteristic lengthscale of the irregularities
and explains the frequent observation of subdiffusion in self-similar fractals.

Several approaches to study anomalous diffusion are based on direct solutions of
the transport problems inside structures that represent those media under certain approximations.
Many of these structures are deterministic fractals \cite{mandelbrot}, which are generated by
recursive application of a rule for generation of porous and solid phases.
Well known examples are the Sierpinski carpets (SCs), whose construction
is illustrated in Fig. \ref{constructionfractals}a,b.
Their relatively simple geometry (e. g. if compared with stochastically generated fractals)
facilitates quantitative or qualitative connections between structural and transport properties.
Random walks were already intensively studied in fractal lattices with the geometry of SCs,
with finite or infinite ramification, and in randomized versions of the SCs
\cite{benavraham1983,kim1993,fssrw,dasgupta,rwpla,rrf,kim2007,babJCP,barlow,
haber2013,darazs,balankin2015,suwannasen}.

In infiltration of a fluid or a solute in a porous medium, an external surface
is in contact with a reservoir of the species that is transported in the pores.
These features are observed in a large variety of systems, such as hydration of rocks, water or
dye absorption in soils or rocks, injection of liquids in fractures or nanoporous solids, etc.
In some cases, models of convective/advective motion and diffusion are studied in deterministic
or randomized fractals
\cite{kuntz,gerolymatou,stalgorova,atzeni,roubinet,martys,perfect2006,dentz2006}.
However, in several other cases, diffusion is the dominant mechanism in the infiltration problem,
which motivates the study of anomalous diffusion models and the study of the
geometry of porous or fractured media \cite{persson,lockington,xu2006,bru,
atzeni,gerasimov,voller,filipovitch,gisladottir}.
The infiltration of randomly moving particles in planar and three-dimensional lattices was also
illustrated in Ref. \protect\cite{sapovalbook}; this motivated the gradient
percolation problem, in which two lattice borders were kept with fixed concentrations of particles
\cite{sapovaloriginal,bunde,rosso1986}.

Fluid infiltration in fractals was considered in a recent work by Voller \cite{voller},
who studied the diffusive motion inside several SCs keeping one external border with
constant pressure.
The fraction of the area occupied by the fluid, which here we call the filling $F$, scales as
\begin{equation}
F \sim t^n ,
\label{F}
\end{equation}
with $n<1/2$, consistently with anomalous subdiffusion.
Subsequently, a Hele-Shaw cell was designed by Filipovitch et al \cite{filipovitch} to reproduce
the pore-block geometry of the SCs and used to study infiltration of glycerin.
The exponents $n$ measured in the experimental apparatus were consistent with the previous
simulation values, thus providing a clear macroscopic demonstration of the relation between
structural disorder and subdiffusion.
Hereafter, these processes are called diffusive infiltration.

In two- or three-dimensional unobstructed lattices, $n$ has the normal diffusion value $1/2$.
However, the exponents $n$ and $\nu$ are very different in the same fractal.
For instance, in the fractal in Fig. \ref{constructionfractals}a, simulation of random walks (RWs)
give $\nu\approx 0.476$ \cite{kim1993,fssrw,suwannasen},
while the infiltration simulations give $n=0.419$ \cite{voller} and the corresponding
experiments give $n=0.423$ \cite{filipovitch}.
Although the works on diffusive infiltration in SCs consider only their first three or four stages
of construction, the finite sizes seem to have small effects of $n$.

The main aim of the present work is to relate the anomalous exponents of single particle diffusion
in the bulk ($\nu$) and of the diffusive infiltration from the border ($n$) of deterministic fractals.
We consider SCs, whose dimensions are between $1$ and $2$, and Menger sponges (MSs), whose
dimensions are between $2$ and $3$.
Numerical simulations are used to obtain accurate estimates of $\nu$ and $n$ and a scaling
approach is used to show that their ratio depends only on the bulk fractal dimension and on the
fractal dimension of the boundary from which the particles come.
We also define a diffusion front in this infiltration problem, show that averaged fronts
in SCs have shapes similar to those of the experiments in the Hele-Shaw cells, and briefly discuss
their roughening in the SCs and MSs.
We stress that our work is concerned with unbiased RWs for both problems, thus it is
not expected to describe systems in which convective or advective transport is relevant.

This paper is organized as follows.
Sec. \ref{model} presents the models of fractal lattices, diffusion processes, and information on the
simulation work.
Sec. \ref{simulations} shows simulation results for single particle diffusion and diffusive infiltration
in SCs and MSs.
Sec. \ref{scaling} presents an approach to connect the scaling exponents of those problems.
In Sec. \ref{fronts}, the roughening of the diffusion fronts is analyzed.
In Sec. \ref{conclusion}, our results and conclusions are summarized.

\section{Fractal lattices, diffusion models, and their simulation}
\label{model}

\begin{figure}[!h]
\includegraphics[width=\textwidth]{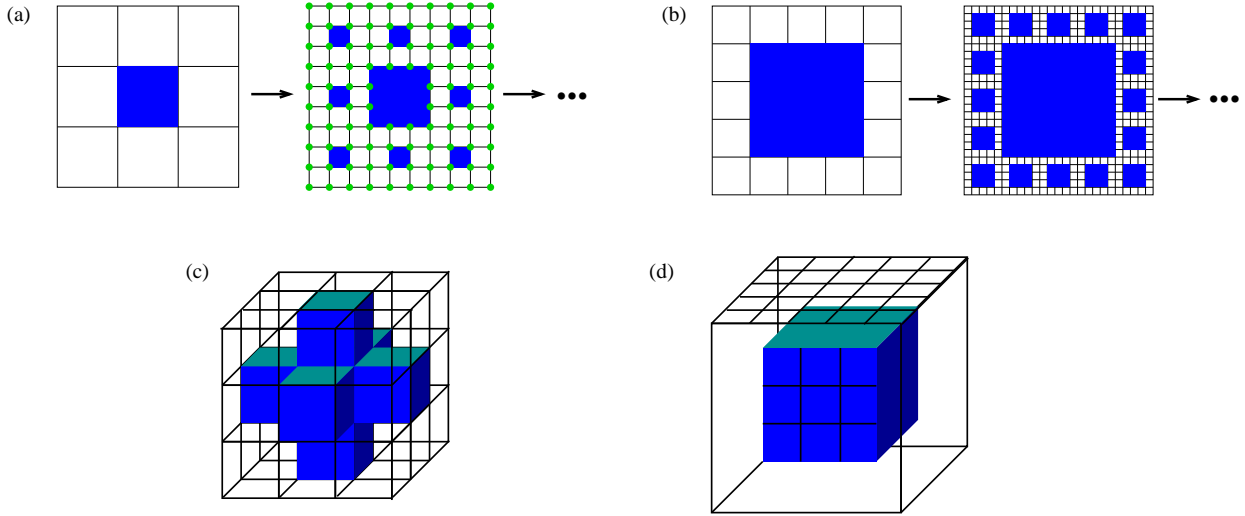}
\caption{(Color online)
Panels (a) and (b) show the first two stages of construction of SC1 and SC2, respectively.
The generator (stage $m=1$) is a square divided in $b^2$ subsquares with $k$ of them removed (blue):
In SC1, $b=3$ and $k=1$; in SC2, $b=5$ and $k=9$.
In each stage of the construction, each remaining square is replaced by the generator, thus $b$ is
the scaling factor of this process.
A SC with dimension $D_F=\ln{\left( b^2-k\right)}/\ln{b}$ is obtained after infinite iterations.
In panel (a), the pore sites of the SC1 lattice are shown as green dots in the stage $m=2$.
Panels (c) and (d) show the generators of MS1 and MS2, respectively, in which a cube is divided in $b^3$
subcubes and $k$ of them (in blue tones) are removed: in MS1, $b=3$ and $k=7$; in MS2, $b=5$ and $k=27$,
but the division of this generator is shown only in one external face of the main cube and one face
of the removed cube to facilitate visualization.
In each stage of the construction of a MS, each remaining cube is replaced by the generator, so that
a fractal with dimension $D_F=\ln{\left( b^3-k\right)}/\ln{b}$ is obtained after infinite iterations.
}
\label{constructionfractals}
\end{figure}

The construction of the SCs studied here is shown in Figs. \ref{constructionfractals}a and
\ref{constructionfractals}b; they are respectively called SC1 and SC2.
Their fractal dimensions are $D_F^{\left( 1\right)}=\ln{8}/\ln{3}$ and
$D_F^{\left( 2\right)}=\ln{16}/\ln{5}$, respectively.
These values up to five decimal places are shown in Table \ref{tableresults}.

A lattice is defined with sites at the vertices of the squares produced at each step
of the construction of the SC.
In each stage $m$, the unit size is defined as the distance between nearest neighbor
sites, thus the lateral size of the lattice (number of sites in one border) is $L=b^m+1$,
where $b$ is the scaling factor of the generator.

The solid sites of the lattice are those located inside the lacunas.
The remaining sites form the pore network. 
The distribution of pore sites is illustrated in the stage $m=2$ of SC1 in
Fig. \ref{constructionfractals}a.
Note that many pore sites are in the borders of the lacunas.
Hereafter we refer to this pore network as the SC; it actually has the same fractal dimension
of the region remaining after infinite iterations of the construction rule.
The particles executing RWs in the SC can occupy only pore sites.

\begin{figure}[!h]
\includegraphics[width=0.4\textwidth]{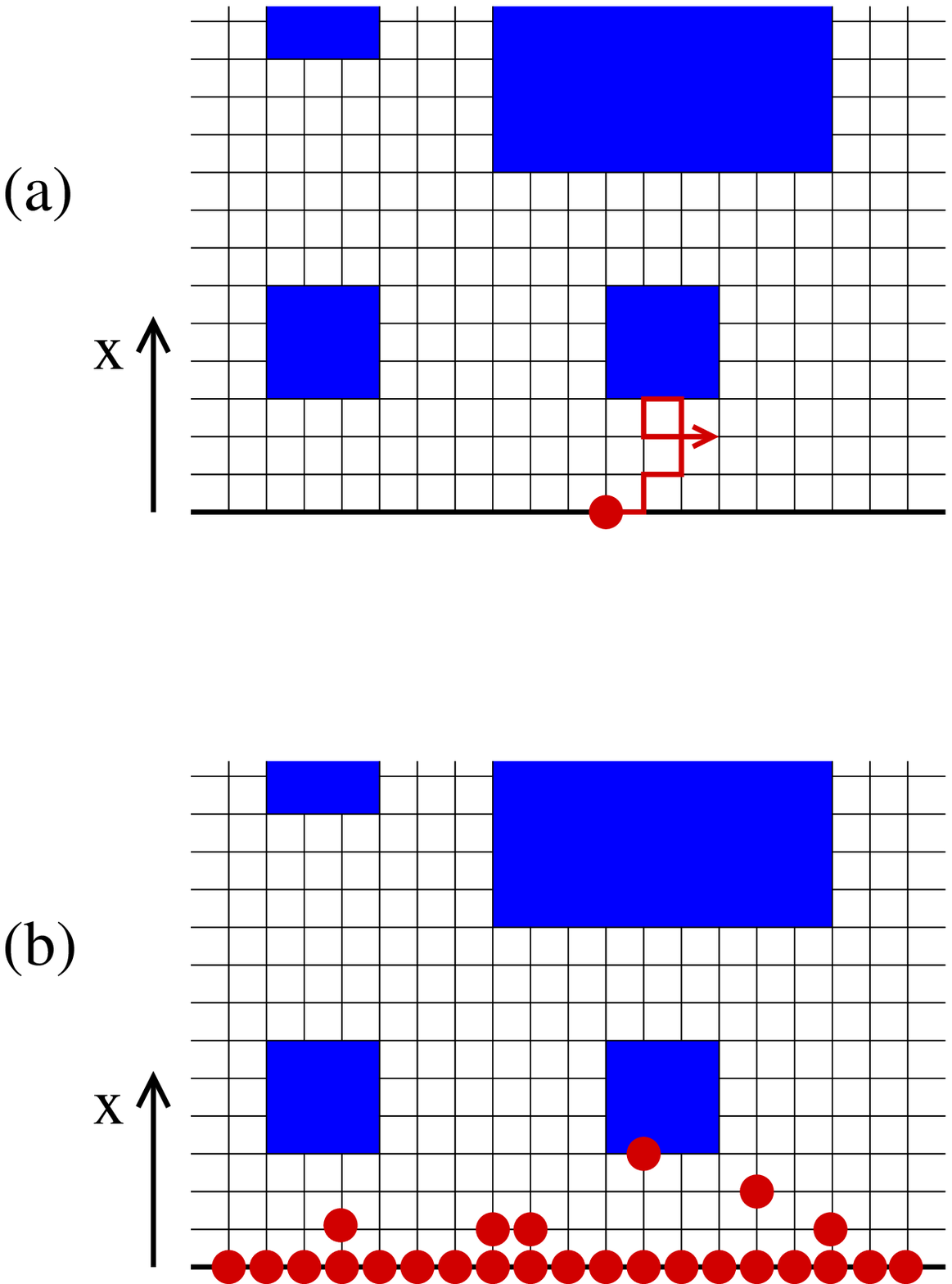}
\caption{(Color online) (a) Some steps of a particle (red) in the single particle diffusion model
in SC1 (white pores, blue solid).
The lower dark line is the impenetrable border ($x=0$).
(b) Configuration of the diffusive infiltration model in SC1 after some steps, with particles filling
the border $x=0$.
}
\label{diffmodel}
\end{figure}

An impenetrable border of the lattice is located at the $y$ axis ($x=0$), as shown in Fig.
\ref{diffmodel}a.
This means that no particle can jump to points with $x<0$.
Periodic boundary conditions are considered in the $y$ direction.
These conditions do not affect the geometric properties of the fractals.

The first step of our work is to study single particle infiltration in the SCs,
with starting positions randomly chosen in the $y$ axis.
This is equivalent to the infiltration of non-interacting particles starting at that axis at $t=0$,
as proposed in a recent model of diffusion in porous deposits \cite{diffballistic}.
In one time unit, the particle randomly chooses one nearest neighbor site to jump to, and moves
to that site only if it is also a pore site; otherwise, the particle does not move.
Fig. \ref{diffmodel}a illustrates the first steps of a particle in SC1.

We simulated ${10}^7$ single particle RWs in the stages $m=6$ to $m=9$ of SC1 and $m=7$ of SC2.
The maximal time of each walk was $t_{MAX}={10}^7$ in the largest lattices.
These conditions ensure that no walker reaches the border at $x=L$.

The model of diffusive infiltration in the SCs is defined analogously to the model in planar
and cubic lattices shown in Ref. \protect\cite{sapovalbook}.
The $y$ axis (line $x=0$) is permanently filled with mobile particles that
execute RWs with excluded volume interactions, i. e. with at most one particle per site.
In one time unit, each particle executes an average of one step trial to a randomly chosen nearest
neighbor site.
The step is allowed if the target site is a pore site and is not occupied by another diffusing
particle; otherwise, the particle does not move.
If a particle leaves the $y$ axis, another particle immediately refills the available position.
This creates a pressure for the particles to move to the positive $x$ direction.
Fig. \ref{diffmodel}b illustrates the beginning of this process in SC1.

In our simulations, $50$ independent configurations of diffusive infiltration were generated
in stages $m=7$ of SC1 and $m=5$ of SC2, with maximal times $t_{MAX}={10}^5$.
Simulations in $m=6$ of SC1 were also performed to confirm the absence of finite-size effects.

The generators of the MSs studied here are shown in Figs. \ref{constructionfractals}c and
\ref{constructionfractals}d; these fractals are respectively called MS1 and MS2.
Those images differ from usual presentations of these fractals because they highlight the solid
region (dark) of the generator, with the remaining region being the porous one.
The fractal dimensions of the porous regions are
$D_F^{\left( 1\right)}=\ln{20}/\ln{3}$ for MS1 and
$D_F^{\left( 2\right)}=\ln{98}/\ln{5}$ for MS2.
The values up to five decimal places are also shown in Table \ref{tableresults}.

\begin{table}
\centering
\caption{Bulk and border dimensions of each fractal, best estimates of exponents, and
corresponding estimates of $\nu\left( D_F-D_B\right)$ for the test of Eq. (\ref{nnu}).
Simulation data were obtained in this work except where indicated.}
\label{tableresults}
\begin{tabular}{|l|l|l|l|l|l|}
\hline
Fractal & $D_F$     & $D_B$     & $\nu$ (simulation)                      & $n$ (simulation) & $\nu\left( D_F-D_B\right)$ \\ \hline
SC1     & $1.89279$ & $1$       & $0.475\pm 0.003$ (Ref. \protect\cite{kim1993}) & $0.424\pm 0.004$ & $0.424\pm 0.003$           \\ \hline
SC2     & $1.72271$ & $1$       & $0.455\pm 0.003$ (Ref. \protect\cite{kim1993}) & $0.334\pm 0.014$ & $0.329\pm 0.002$           \\ \hline
MS1     & $2.72683$ & $1.89279$ & $0.467\pm 0.005$ & $0.389\pm 0.002$ & $0.389\pm 0.004$           \\  \hline
MS2     & $2.84880$ & $2$       & $0.479\pm 0.013$ & $0.407\pm 0.014$ & $0.407\pm 0.012$        \\  \hline
\end{tabular}
\end{table}

Lattice sites are located at the vertices of the cubes produced at each step of the construction
of the MS and the distance between nearest neighbor sites is taken as the size unit.
At stage $m$, the lateral size of the lattice is $L=b^m+1$, where
$b$ is the scaling factor of the generator.
The solid sites are located inside the lacunas at each stage and the remaining sites are pore sites,
which may be occupied by particles executing RWs.
An impenetrable border of the MS lattice is located at the $yz$ plane ($x=0$) and
periodic conditions are considered in the directions $y$ and $z$.

In single particle infiltration in MSs, each particle is released at a randomly chosen
pore site of the border $x=0$ ($yz$ plane) and, in each time unit, chooses one nearest neighbor site
and jumps to that site only if it is also a pore site.
${10}^6$ RWs were generated in stages $m=6$ of MS1 and $m=4$ of MS2, with $t_{MAX}={10}^5$.

In diffusive infiltration in MSs, all pore sites of the border $x=0$ are permanently occupied
by mobile particles and each particle executes an average of one step trial per unit time;
when a pore site at $x=0$ becomes empty, it is instantaneously refilled.
In stages $m=6$ of MS1 and $m=4$ of MS2, we produced $20$ configurations of diffusive infiltration,
with maximal times $t_{MAX}={10}^4$.

A diffusion front $\{ h\}$ is defined in the diffusive infiltration problem.
The front height at position $y$ of a SC [$\left( y,z\right)$ of a MS] is an average
of the displacements $x$ of all particles with that position.
In general, at a given substrate position $i$, the front height $h_i$ is 
\begin{equation}
h_i \left( t \right)\equiv \frac{2}{N_i}\sum_{\sigma =1}^{\sigma =N_i}{x_\sigma} ,
\label{defh}
\end{equation}
where $N_i$ is the number of particles with that substrate position and $\sigma$ runs
over all those particles.
If this definition is used for a configuration with no vacancy between particles (solid-on-solid
aggregates), then $h_i$ is equal to the position $x$ of the top particle at position $i$;
this is the usual definition of the interface in film growth and/or kinetic roughening models,
and justifies the factor $2$ in Eq. \ref{defh}.

The roughness of the diffusion front, $W\left( t\right)$, was calculated for selected times.
It is defined as the rms fluctuation of $\{ h\}$, averaged over the substrate positions
and over different configurations of the front at time $t$.
Finite-size effects on $W$ are expected only when $\langle h\rangle \sim L^z$ or longer,
where $z>1$ is the dynamical exponent of the front roughening \cite{barabasi}.
However, the infiltration simulations are restricted to $\langle h\rangle<L$, thus 
$W$ is not expected to depend on $L$, i. e. roughening is in the growth regime \cite{barabasi}.

\section{Simulation results}
\label{simulations}

\subsection{Infiltration in Sierpinski carpets}
\label{simulationSC}

\begin{figure}[!h]
\includegraphics[width=0.5\textwidth]{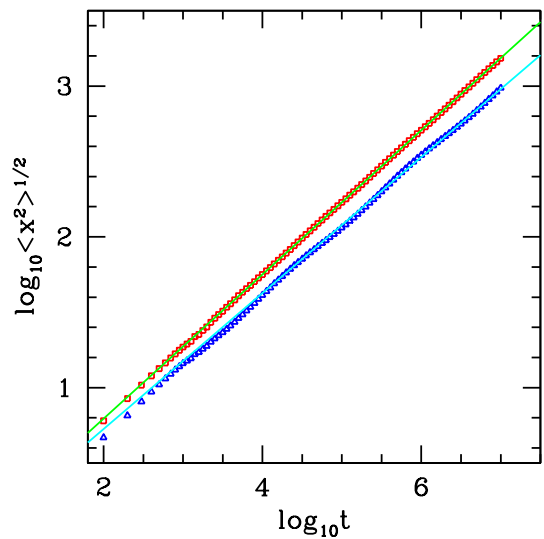}
\caption{(Color online) Mean square displacement as a function of time of single particle RWs in
stage $m=9$ of SC1 (red squares) and $m=7$ of SC2 (blue triangles).
For clarity, we show only data points in intervals $\geq 0.05$ of $\log_{10}t$.
Solid lines are least squares fits of all the data generated in the range ${10}^3\leq t\leq{10}^7$.
}
\label{x2SC}
\end{figure}

Fig. \ref{x2SC} shows the time evolution of the mean square displacement in the $x$ direction
in single particle diffusion in SC1 ($m=9$) and SC2 ($m=7$).
The linear fits of each data set are shown.

The mean square displacement oscillates in both cases, but the oscillations are visible only
in the data for SC2.
The diffusing particle may take a long time to go around the border of lacunas, whose sizes
increase as the particle travels to more distant points.
The amplitude of oscillations are much smaller in SC1 because its lacunas are relatively small
if compared to those of SC2.
These oscillations are log-periodic, similarly to those observed in simulations of RWs in the
bulk of SCs \cite{babJCP}, and are consequence of the discrete scale invariance of those
fractals \cite{babEPL,akkermans}.

The estimates $\nu =0.478\pm 0.003$ for SC1 and $\nu=0.45\pm 0.01$ for SC2 are obtained
by performing linear fits in several time ranges in the simulated interval.
Fits of the data in smaller stages of construction of those fractals
confirm that finite-size effects are negligible.

These estimates are very close to those obtained from simulations
of RWs starting at random points of the bulk of the SCs:
$\nu=0.475\pm 0.003$ \cite{kim1993} and $0.476\pm 0.005$ \cite{fssrw,suwannasen} in SC1;
$\nu=0.458\pm 0.004$ \cite{rrf} and $0.455\pm 0.003$ \cite{kim1993} in SC2.
This comparison is important because it shows that the long time properties of single particle
RWs in the SCs do not depend on the initial positions of those particles nor on the boundary
conditions.
We understand that this is consistent with the uniqueness of Brownian motion in a
Sierpinski carpet demonstrated in Ref. \protect\cite{barlow}, which is related to the
uniqueness of the Laplacian definition in that fractal.
The most accurate estimates of $\nu$ for each fractal are shown in Table \ref{tableresults}.

\begin{figure}[!h]
\includegraphics[width=0.6\textwidth]{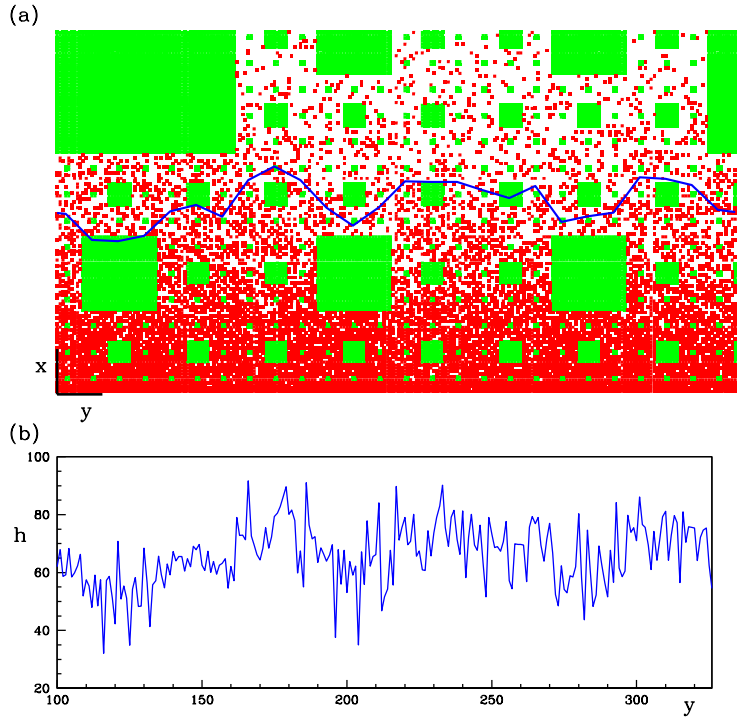}
\caption{(Color online)
(a) Configuration of the diffusive infiltration in SC1 at $t=8000$, with lacunas (solid) in light green,
empty (pore) region in white, and diffusing particles in red.
The blue curve is the averaged diffusion front $\{ H\}$.
(b) Diffusion front (not averaged) $\{ h\}$ in the same positions $y$ of the above picture.
}
\label{infiltrationSC}
\end{figure}

The diffusive infiltration in SC1 is illustrated in Fig. \ref{infiltrationSC}a, which shows a region
near the filled border of that lattice at time $t=8000$.
Fig. \ref{infiltrationSC}a also shows an averaged diffusion front $\{ H\}$, in which
$H\left( 9j\right)$ (position $y=9j$, with integer $j\geq 0$)
is the average of the heights $h_i$ (Eq. \ref{defh}) from $i=9j-4$ to $i=9j+4$.
This is an average over $9$ substrate positions, which is the second smallest size of the lacunas
in the lattice.
This averaging highlights the long wavelength fluctuations.
Instead, the diffusion front $\{ h\}$, which is shown in Fig. \ref{infiltrationSC}b,
also has short wavelength fluctuations of large amplitude.

The averaged diffusion front has a structure of rounded mounds separated by gaps, with the
mounds located between the third level lacunas.
The main depletion of that front is close to the fourth level lacuna (the largest one at the
left side).
This morphology resembles that observed in infiltration of glycerin in the Hele-Shaw cells in
Ref. \protect\cite{filipovitch}, which reinforces the connection with that system.
The experimental front is smoother, but this is probably related
to interfacial tension effects and to the small stage of the SC used in the cell.

\begin{figure}[!h]
\includegraphics[width=0.5\textwidth]{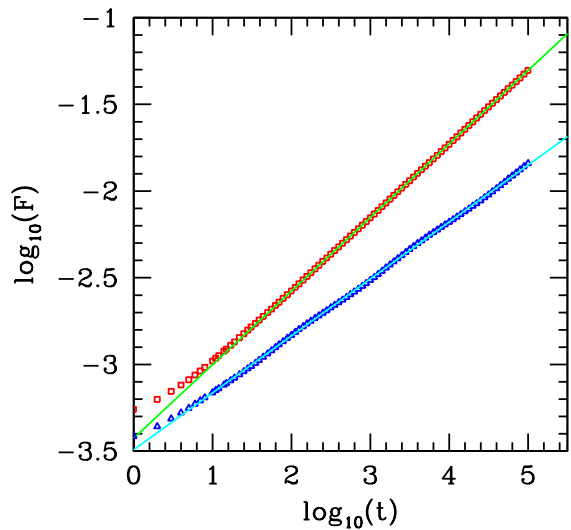}
\caption{(Color online)
Filling of stage $m=7$ of SC1 (red squares) and $m=5$ of SC2 (blue triangles) as a function of time
in the diffusive infiltration model.
Only data points with difference $\geq 0.05$ in $\log_{10}t$ were plotted.
Solid lines are least squares fits of all the data generated in the range ${10}^2\leq t\leq{10}^5$.
}
\label{FSC}
\end{figure}

The filling $F\left( t\right)$ is the number of moving particles at time $t$ per lattice site.
Fig. \ref{FSC} shows the time evolution of $F$ in SC1 ($m=7$) and SC2 ($m=5$) and linear fits
of each data set.
Considering fits in various time ranges, we obtain the estimates $n=0.424\pm 0.004$ and
$n=0.334\pm 0.014$, respectively, which are reproduced in Table \ref{tableresults}.

The value of $n$ in SC1 is very close to the estimate $n=0.419$ of Ref. \protect\cite{voller} for
infiltration simulated with a diffusion equation; in SC2, $n=0.319$ was obtained in that work,
which differs $4.5\%$ from our estimate.
The experiments of glycerin infiltration in the Hele-Shaw cells of
Ref. \protect\cite{filipovitch} give $n=0.423$ and $n=0.334$, respectively, which are
both in excellent agreement with our results.
These results suggest a universal scaling in the diffusive infiltration problem in SCs.

The exponents $n$ are much smaller than the estimates of $\nu$ in the same SCs.
The differences are $-11.5\%$ for SC1 and $-27.1\%$ for SC2, both much larger than error bars
of the estimates of both exponents.

We also performed simulations of single particle RWs and of diffusive infiltration in free
square lattices, i. e. without obstacles.
Even with a small number of configurations in a lattice of lateral size $1024$ and with
maximal simulation time ${10}^5$, we obtained $n\approx \nu\approx 0.5$ with good accuracy.

\subsection{Infiltration in Menger sponges}
\label{simulationMS}

\begin{figure}[!h]
\includegraphics[width=0.5\textwidth]{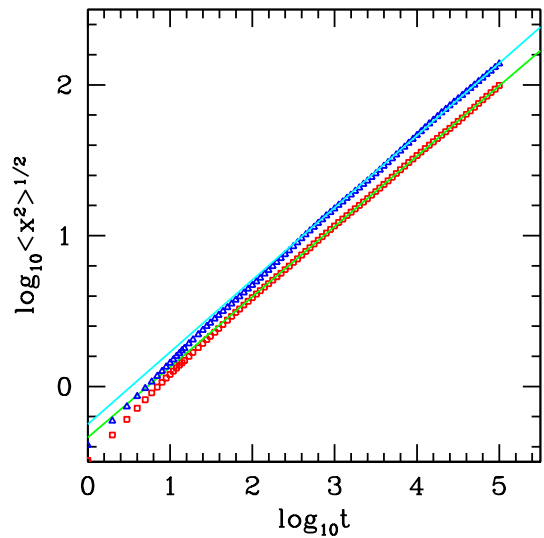}
\caption{(Color online)
Mean square displacement as a function of time of single particle RWs in
stage $m=6$ of MS1 (red squares) and $m=4$ of MS2 (blue triangles).
Only data points with differences $\geq 0.05$ in $\log_{10}t$ were plotted.
Solid lines are least squares fits of all the data generated in the range ${10}^2\leq t\leq{10}^5$.
}
\label{x2MS}
\end{figure}

Fig. \ref{x2MS} shows the time evolution of the mean square displacement in the $x$ direction
of single particle diffusion in MS1 ($m=6$) and MS2 ($m=4$), with linear fits of each data set.
The log-periodic oscillations due to the discrete scale invariance \cite{babJCP,babEPL} are
also observed here, and their amplitudes are also larger in the fractal with larger lacunas (MS2).

\begin{figure}[!h]
\includegraphics[width=0.7\textwidth]{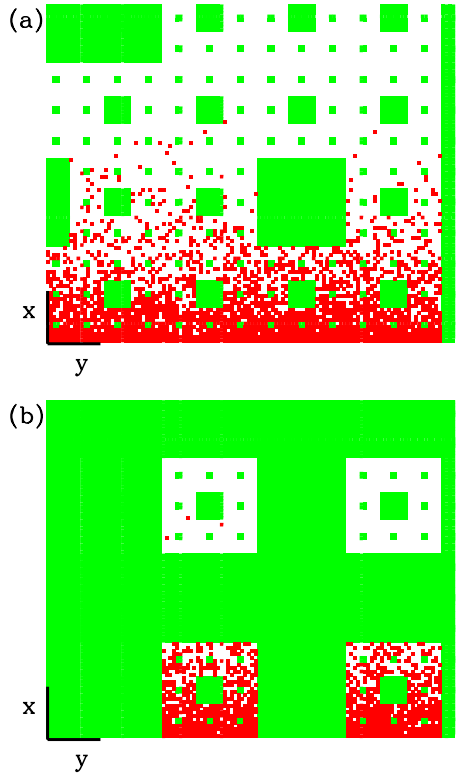}
\caption{(Color online)
Configurations of diffusive infiltration at $t=2000$ in planes (a) $z=250$ and (b) $z=360$ of MS1.
Colors are the same as in Fig. \ref{infiltrationSC}.
}
\label{infiltrationMS}
\end{figure}

Fits of the data in various time intervals yield the estimates $\nu =0.467\pm 0.005$ for MS1 and
$\nu=0.479\pm 0.013$ for MS2, which are also reproduced in Table \ref{tableresults}.
Previous estimates of $\nu$ in MSs were obtained only from lower and upper bounds
\cite{benavraham1983,balankin2015}, thus they had lower accuracy than the present ones.
Simulations in smaller stages of MS1 and MS2 give approximately the same estimates, indicating that
finite-size effects are small.

Note that the exponent $\nu$ in the MSs follows the same trend of decrease with the fractal
dimension that was observed in the SCs in Ref. \protect\cite{fssrw}.
Moreover, those exponents are near the normal diffusion value $1/2$, which is also observed in
fractals without dead ends and dimensions between $1$ and $2$, such as the SCs \cite{rrf}.

The diffusive infiltration is illustrated in Figs. \ref{infiltrationMS}a,b, which
show cross sections of MS1 near the filled boundary at $t=2000$.
In the plane $z=250$ (Fig. \ref{infiltrationMS}a), the density of obstacles near the filled
boundary is low, thus it is easier for particles to reach larger distances.
In the plane $z=360$ (Fig. \ref{infiltrationMS}b), the density of blocks near the boundary is
large, which confines many particles; however, note that three particles have already reached
an upper porous region of this plane by migrating through other planes.

\begin{figure}[!h]
\includegraphics[width=0.5\textwidth]{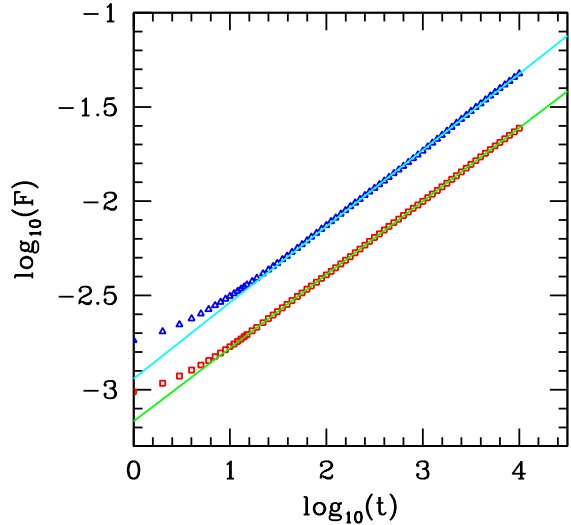}
\caption{(Color online)
Filling of stage $m=6$ of MS1 (red squares) and $m=4$ of MS2 (blue triangles) as a function of time
in the diffusive infiltration model.
Only data points with differences $\geq 0.05$ in $\log_{10}t$ were plotted.
Solid lines are least squares fits of all the data generated in the range ${10}^2\leq t\leq{10}^5$.
}
\label{FMS}
\end{figure}

Fig. \ref{FMS} shows the time evolution of the filling $F$ in MS1 ($m=6$) and MS2 ($m=4$).
Linear fits of the data with ${10}^2\leq t\leq {10}^4$ are shown.
We also analyzed fits in different time ranges to obtain the estimates
$n=0.389\pm 0.002$ and $n=0.407\pm 0.014$, respectively.
They are presented in Table \ref{tableresults}.
Again, we also observe that the exponents $\nu$ and $n$ are very different.

We also simulated single particle RWs and the diffusive infiltration model in simple cubic
lattices.
With a small number of configurations and maximal time ${10}^5$, we obtained
$n\approx \nu\approx 0.5$, which is consistent with normal diffusion in both cases.

\section{Scaling approach}
\label{scaling}

\begin{figure}[!h]
\includegraphics[width=0.5\textwidth]{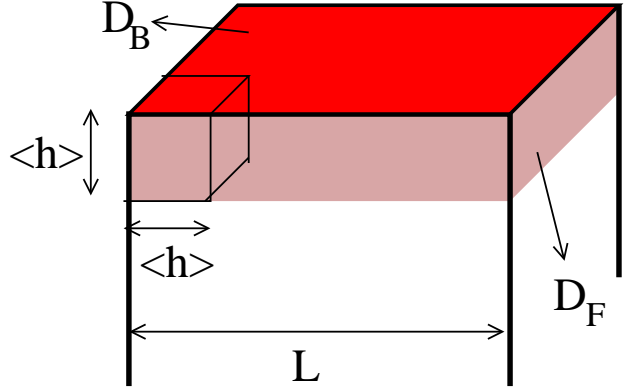}
\caption{(Color online)
Scheme of diffusive infiltration from a border characterized by fractal dimension $D_B$
to a medium characterized by fractal dimension $D_F$ with lateral size $L$.
$\langle h\rangle$ is the average thickness of the diffusion front.
}
\label{scheme}
\end{figure}

A scheme of the diffusive infiltration in a fractal is shown in Fig. \ref{scheme},
with a characteristic length $\langle h\rangle$ filled by the diffusing species.
$L$ is the lateral size of the lattice, whose dimension is $D_F$ and whose filled boundary
has dimension $D_B$.
For SCs, the boundary is a filled line, thus $D_B=1$; for MS2, the boundary is a plane,
thus $D_B=2$; however, for MS1, the filled boundary has the geometry of SC1, thus it
has dimension $D_B=\ln{8}/\ln{3}\approx 1.89279$.

The diffusing front is expected to advance with the same scaling of single particle diffusion
because the more advanced particles move in a region with low density, in which the main
constraints are the irregularities of the fractal network and not the excluded volume
interactions.
For this reason, we expect
\begin{equation}
\langle h\rangle \sim t^\nu .
\label{hmedio}
\end{equation}
The value of $\langle h\rangle$ calculated in our simulations are consistent with this scaling,
but fluctuations are much larger than those of single particle diffusion.

The filled region can be divided in hypercubes of edge $\langle h\rangle$; one of them is highlighted
in Fig. \ref{scheme}.
The total filling of each hypercube is
\begin{equation}
F_H\sim \langle h\rangle^{D_F} ,
\label{FH}
\end{equation}
assuming that $\langle h\rangle$ is sufficiently large for the fractality of the medium
to be observed.
If $L\gg \langle h\rangle$, the number of hypercubes in the boundary is
\begin{equation}
N_H\sim {\left( \frac{L}{\langle h\rangle}\right)}^{D_B} .
\label{NH}
\end{equation}

The total filling consequently scales as
\begin{equation}
F = N_H F_H\sim L^{D_B}{\langle h\rangle}^{D_F-D_B} \sim L^{D_B} t^{\nu\left( D_F-D_B\right)} .
\label{Fscaling}
\end{equation}
This gives an exact relation between the single particle diffusion exponent and the diffusive
infiltration exponent:
\begin{equation}
n = \nu\left( D_F-D_B\right) = \frac{D_F-D_B}{D_W} ,
\label{nnu}
\end{equation}
where Eq. (\ref{DW}) was also used.

Table \ref{tableresults} shows the values of $\nu\left( D_F-D_B\right)$ obtained from the best
available estimates of $\nu$ and the exact dimensions $D_F$ and $D_B$.
They are in excellent agreement with the estimates of $n$ obtained in simulations.
Note that Eq. \ref{nnu} implies $n=\nu=1/2$ for systems in which bulk and boundary are regular
lattices with integer dimensions, since $D_B=D_F-1$ and single particle diffusion is normal in
those cases.
Also note that the successful application to systems whose boundaries are
compact (SCs and MS2) and fractal (MS1) is a strong support to this scaling approach.

In Ref. \protect\cite{fssrw}, it was shown that the exponent $\nu$ in SCs has an approximately
linear dependence on $D_F$ if this dimension is not much smaller than $2$.
Combination of this relation with Eq. (\ref{nnu}) gives an approximately quadratic dependence
of $n$ on $D_F$.
Indeed, such a quadratic relation was obtained by Filipovitch et al \cite{filipovitch} in the
experiments of fluid infiltration in SCs.
For many other fractals in which the RW exponent is known, Eq. \ref{nnu} can predict the
anomalous properties of diffusive infiltration and, if experiments are available, it may
help to evaluate the applicability of a given fractal model.

\section{Roughening of the infiltration fronts}
\label{fronts}

\begin{figure}[!h]
\includegraphics[width=0.8\textwidth]{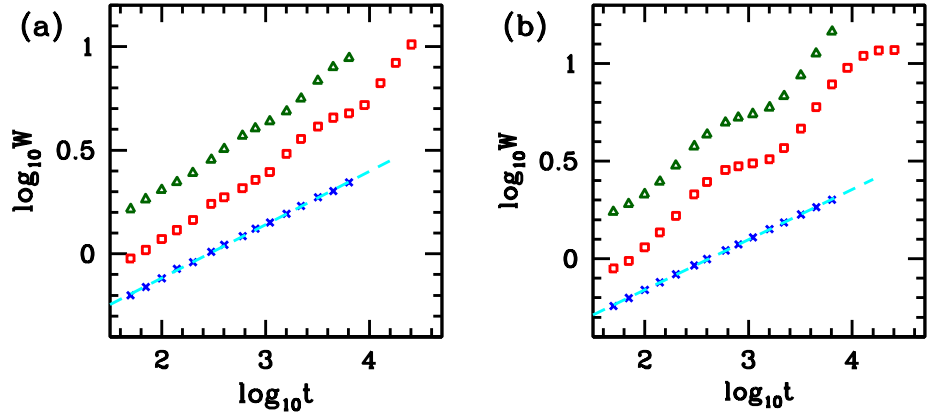}
\caption{(Color online)
Roughness of diffusion fronts in: (a) SC1 (red squares), MS1 (green triangles), and square lattice (blue
crosses); (b) SC2 (red squares), MS2 (green triangles), and simple cubic lattice (blue crosses).
The data in the square (cubic) lattice is displaced $0.2$ ($0.5$) units to the bottom to avoid
intersection with other data sets.
Dashed lines are least squares fits of data in square and cubic lattices, 
both with slope $0.257$.
}
\label{roughness}
\end{figure}

The roughness of the diffusion front was measured in all fractals at selected times, from
$t=50$ to $t=36000$ in SCs and from $t=50$ to $t=6400$ in MSs.
Fig. \ref{roughness}a shows $W$ as a function of $t$ in SC1, MS1, and in the square lattice;
Fig. \ref{roughness}b shows the same quantities in SC2, MS2, and in the simple cubic lattice.

In square and cubic lattices, the linear fits of the data shown in Figs. \ref{roughness}a,b
give $W\sim t^\beta$ with a growth exponent $\beta\approx 0.25$.
The movement of the particles in the diffusion front is completely random,
thus the height of each position $y$ in the SCs ($yz$ in MSs) randomly fluctuates around
the average value $\langle h\rangle$.
This is characteristic of random uncorrelated deposition, in which
$W\sim \langle h\rangle^{1/2}$ \cite{barabasi}.
Since $\langle h\rangle\sim t^{1/2}$ in diffusive infiltration in those lattices,
we obtain $\beta=1/4$, which is consistent with the simulation results.

The infiltration problem defined here has similarities with that of gradient percolation,
in particular the existence of lattice borders with fixed concentration of particles;
see e. g. Ref. \protect\cite{sapovalbook}.
However, a very important difference is the existence of a fixed concentration gradient
along the $x$ direction in that case; instead, in the present problem, the concentration
gradient is continuously varying between the filled border and the diffusion front.
In the gradient percolation problem, the diffusion front is defined as the interface of
a cluster of connected particles, which also differs from the present definition.
Thus, even if correlations in particle positions were introduced in our model
(e. g. to represent surface tension effects), the roughening might be different from
that of the gradient percolation front \cite{bunde,rosso1986}.

Despite the simplicity of the diffusive infiltration front in regular lattices and the
fact that the uncorrelated growth extends to the fractal media, some interesting features
can be observed in the latter case.
As shown in Figs. \ref{roughness}a,b, the roughness oscillates in all fractals.
As the front reaches the lower borders of a set of parallel lacunas of a given size, the front can
advance only in the regions between those lacunas, which leads to large differences in the heights
at the confined and at the non-confined regions; see e. g. Figs. \ref{infiltrationSC}a and
\ref{infiltrationMS}a,b.
However, when the front reaches the upper borders of those lacunas, it enters a more homogeneous
region, in which lateral diffusion slows down the increase of height differences.

\begin{figure}[!h]
\includegraphics[width=0.5\textwidth]{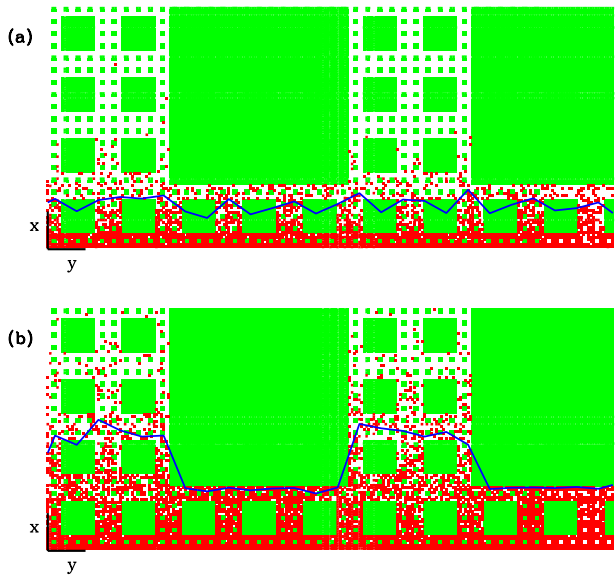}
\caption{(Color online)
Configurations of diffusive infiltration at (a) $t=1000$ and (b) $t=5000$ in SC2, with the
corresponding averaged fronts.
Colors are the same as in Fig. \ref{infiltrationSC}.
}
\label{infiltrationSC53}
\end{figure}

This effect is enhanced in lattices with large lacunas, which is the case of SC2 and MS2.
For instance, Figs. \ref{infiltrationSC53}a shows an infiltration profile in SC2 at
$t=1000$, in which the front has bypassed the second level lacunas but did not reach the
larger ones.
Correspondingly, Fig. \ref{roughness}b shows a plateau in the $\log{W}\times \log{t}$
plot at $t\sim 1000$.
On the other hand, Fig. \ref{infiltrationSC53}b shows an infiltration profile at 
$t=5000$, in which particles enter the gaps between the third level lacunas.
Correspondingly, Fig. \ref{roughness}b shows a rapid increase of $W$ at $t\sim 5000$.
The main contribution to the roughness is that from the long wavelength fluctuations,
which are the height differences between the evolving regions (in the gaps) and the
blocked regions (below the large lacunas).
This is not a kinetic roughening feature, but an effect of the channeled geometry of the medium.
For this reason, it is meaningful to estimate a growth exponent $\beta$ in these cases.

The roughness oscillations are also log-periodic, which is related to the discrete scale
invariance of the medium.
An important feature is that the periods have approximately the same value for the fractals
with the same scaling factors in the generators, even having very different $D_F$ and different
embedding dimensions: $b=3$ for SC1 and MS1 (Fig. \ref{roughness}a)
and $b=5$ for SC2 and MS2 (Fig. \ref{roughness}b).
The relevance of the scaling factor of the generator is consistent with the approach
of Ref. \protect\cite{babEPL} to explain these oscillations in kinetic models.

Other growth models have also been studied in substrates with the geometry of SCs
\cite{horowitz,lee,tang2010,xun2014}.
The roughness oscillations were also observed in the simulations of ballistic deposition
in SCs \cite{horowitz}.
However, a comparison with our results is not possible because the front kinetics and the
substrate effects are very different.
For instance, here the front grows in the plane in which the SC is embedded, thus it finds
different disordered environments in the lateral directions during the growth,
while those works consider growth parallel to the SC plane, so that the lateral disorder
is the same for the growing columns at all heights.

\section{Conclusion}
\label{conclusion}

Although diffusion in deterministic fractals have been intensively studied since a long time,
as reviewed in Refs. \cite{havlin,metzler2014}, novel interesting features and applications
frequently appear, as shown in recent works, e. g. Refs.
\protect\cite{balankin2015,haber2013,darazs,akkermans,forte,miyaguchi,sokolov}.
The simulation of infiltration of a diffusing fluid in a Sierpinski carpet and subsequent
experimental realization of this process in a Hele-Shaw cell provide a very interesting
macroscopic illustration of that phenomena \cite{voller,filipovitch}.
However, the significant discrepancy between the anomalous exponent of filled area
and the exponent of single particle diffusion in the same fractals was not explained.
The main aim of this work was to fill this gap.

We performed numerical simulations of the infiltration of randomly moving particles from a
permanently filled border in deterministic fractals embedded in dimensions $2$ and $3$,
viz. Sierpinski carpets (SCs) and Menger sponges (MSs).
The exponent $n$ of the time scaling of the infiltrated area/volume was measured and
confirms the accuracy of the previous infiltration simulations \cite{voller}
and experiments \cite{filipovitch}, which were obtained in smaller stages of construction of SCs.
Single particle diffusion starting from the same border was also studied numerically and the
exponent $\nu$ of the mean square displacement scaling was measured.
In SCs, the values of $\nu$ agree with previous estimates in the bulk of those fractals;
in MSs, they improve previous estimates, since they were based only on
lower and upper bounds and had large uncertainties.

A scaling approach is proposed to relate exponents $n$ and $\nu$, considering the fractal
dimensions of the infiltrated region and of the region from which the diffusing particles come. 
The numerical results are in excellent agreement with this approach.
Thus, if the dimensions characterizing the porous medium and its boundaries are known,
then the single particle diffusion exponent $\nu$ (which was calculated for a
variety of fractals in more than three decades) is sufficient to determine
the scaling properties in the diffusive infiltration problem.

We also showed that the roughness of the diffusion fronts has log-periodic oscillations
in time, which is characteristic of random walks and other kinetic models in fractals
\cite{babJCP,babEPL,akkermans}.
The same oscillations are observed in the mean-square displacement of single particle RWs.
In SCs and MSs whose generators have the same scaling factor, the periods are approximately the
same, despite the very different dimensions of those fractals, which shows the relevance of
the discrete scale invariance.

\begin{acknowledgments}
The author thanks Vaughan Voller for helpful discussion.

This work was supported by CNPq and FAPERJ (Brazilian agencies).
\end{acknowledgments}


\vfill\eject

\end{document}